\documentclass{article}
\usepackage{spconf,amsmath,graphicx, booktabs}
\usepackage[table]{xcolor}
\usepackage{amssymb} 
\usepackage[T1]{fontenc}
\usepackage[hidelinks]{hyperref}

\definecolor{mygreen}{rgb}{0.1, 0.6, 0.1}
\definecolor{myred}{rgb}{0.8, 0.2, 0.2}

\newcommand\blfootnote[1]{%
  \begingroup
  \renewcommand\thefootnote{}\footnote{#1}%
  \addtocounter{footnote}{-1}%
  \endgroup
}

\newcommand{\sign}{\ifdim 0pt=0pt \else \fi} 

\newcommand{\perfdelta}[1]{%
  \ifdim #1 pt < -0.5 pt 
    {\color{myred}\ifdim #1 pt > 0pt +\fi #1}%
  \else 
    \ifdim #1 pt < 0.51 pt 
      \ifdim #1 pt > 0pt +\fi #1%
    \else 
      {\color{mygreen}\ifdim #1 pt > 0pt +\fi #1}%
    \fi
  \fi
}

\title{WavLink: Compact Audio-Text Embeddings with a Global Whisper Token}
\name{Gokul Karthik Kumar\quad Ludovick Lepauloux\quad Hakim Hacid}
\address{Technology Innovation Institute, Abu Dhabi, UAE}
\begin{document}
%
\maketitle
\begin{abstract}
Whisper has become the de-facto encoder for extracting general-purpose audio features in large audio–language models, where a 30-second clip is typically represented by 1500 frame features projected into an LLM. In contrast, audio–text embedding models like CLAP-based models have largely relied on alternative audio encoders (e.g., HTS-AT, PaSST), and have not leveraged Whisper effectively. We present WavLink, a compact audio–text embedding model that augments Whisper encoder with a learnable global token, trained jointly with a text encoder. Through a systematic study of design choices, including pretrained text encoders, loss functions, training modes, and data mixtures, we identify configurations that yield state-of-the-art retrieval performance. Our two-stage training recipe across three model sizes, combined with Matryoshka-style supervision, improves scalability, enabling 8× smaller embeddings with minimal performance drop. WavLink also demonstrates competitive performance on AIR-Bench with MCQs and zero-shot classification.
\blfootnote{© 2026 IEEE.  Personal use of this material is permitted. Permission from IEEE must be obtained for all other uses, in any current or future media, including reprinting/republishing this material for advertising or promotional purposes, creating new collective works, for resale or redistribution to servers or lists, or reuse of any copyrighted component of this work in other works.}
\end{abstract}

%
%
\section{Introduction}
\label{sec:intro}

Learning joint audio--text representations has become a central problem in speech, sound, and music understanding. Recent progress has been driven by the release of large-scale audio--text datasets, including AudioCaps~\cite{audiocaps}, Clotho~\cite{clotho}, and VGGSound~\cite{vggsound}, as well as synthetically generated corpora such as Auto-ACD~\cite{autoacd} and AudioSetCaps~\cite{audiosetcaps}. These resources have enabled large-scale training of audio--text embedding models such as LAION-CLAP~\cite{laionclap}, MS-CLAP~\cite{msclap}, MGA-CLAP~\cite{mgaclap}, ReCLAP~\cite{reclap}, and AF-CLAP~\cite{af2}, which achieve strong results in retrieval and zero-shot classification. In parallel, the rise of audio-understanding Large Language Models (audio-LLMs) e.g.\ Qwen2-Audio~\cite{qwen2audio}, Falcon3-Audio~\cite{falcon3audio}, SALMONN~\cite{salmonn}, Audio Flamingo 3~\cite{af3}, and Voxtral~\cite{voxtral} demonstrates the effectiveness of projecting audio features into LLMs for instruction-following on audio inputs.
Despite these advances, there remains a methodological divide between audio-LLMs and embedding models. Audio LLMs typically adopt \emph{Whisper}~\cite{whisper} as their audio encoder: a 30s clip yields 1500 frame-level features, which are then projected into the LLM for instruction following. Whisper has also been adapted for audio tagging in Whisper-AT~\cite{whisperat}.
 By contrast, embedding models---tasked with retrieval, captioning, and zero-shot classification---have almost exclusively relied on specialized encoders such as HTS-AT~\cite{htsat} or PaSST~\cite{passt}. Thus, Whisper has become ubiquitous for LLM-based audio reasoning, but has been largely unused for compact audio--text embeddings which motivates our work.

We introduce \textbf{\textit{WavLink}}, a compact audio--text embedding model that leverages Whisper. We augment Whisper with a \emph{learnable global token}, trained jointly with a text encoder. Instead of 1500 frame-level tokens, \textit{WavLink} produces a single representation, yielding savings in storage and similarity search cost. Through design sweeps, we compare CLIP~\cite{clip} and ModernBERT~\cite{modernbert} text encoders, CLIP versus SigLIP~\cite{siglip} losses, and different adaptation regimes (projector-only, LoRA~\cite{lora}, full finetuning), across both audio-only and joint tower updates. The best configuration is then scaled in a two-stage training recipe, with Matryoshka~\cite{matry} supervision for multi-resolution embeddings.

We evaluate \textit{WavLink} across \textbf{retrieval} with AudioCaps and Clotho, \textbf{zero-shot classification} with VGGSound, ESC-50~\cite{esc50} and US8K~\cite{us8k}, and \textbf{MCQ} using AIR-Bench~\cite{airbench}. Results show that \textit{WavLink} not only surpasses prior CLAP variants in retrieval, but also achieves competitive accuracy on multiple-choice QA benchmark compared to much larger audio-LLMs such as Qwen2-Audio and Falcon3-Audio. To the best of our knowledge, \textit{WavLink} is the first to show that sub-100 dimensional embeddings, enabled by Matryoshka supervision, can retain competitive performance. This demonstrates that Whisper’s ASR-pretrained features can be adapted effectively for general audio--text embeddings, and that a single global token can bridge the efficiency gap between frame-level audio-LLMs and compact embedding models.

\begin{table}[t]
  \centering
  \vspace{-6pt}
  \caption{WavLink model specifications.}
  \label{tab:model_specs}
  \small 
  \begin{tabular}{l l l}
    \toprule
    \textbf{Size} & \textbf{Parameters (Audio + Text)} & \textbf{Supported Dimensions} \\
    \midrule
    Large  & 761M (637 + 123) & 768, 384, 192, 96 \\
    Small  & 152M (88 + 63)   & 512, 256, 128, 64 \\
    Base   & 84M (20 + 63)    & 512, 256, 128, 64 \\
    \bottomrule
  \end{tabular}
\end{table}

\clearpage
\section{Method}
\label{sec:method}

\subsection{Model Architecture}
\label{sec:method-arch}

Our goal is to obtain a compact audio embedding from Whisper, which otherwise produces
$\sim$1500 frame tokens for a 30-second clip. We augment Whisper’s encoder with a
\textbf{learnable \textit{global token}} that serves as a content-adaptive aggregator.
Given log-Mel features $X \in \mathbb{R}^{B \times F \times T}$,
the convolutional front-end produces hidden states
$\tilde H_0 \in \mathbb{R}^{B \times T' \times D}$.
We append a parameter vector $\mathbf{a}_{\text{cls}} \in \mathbb{R}^{1 \times D}$
to each sequence and propagate the extended sequence through Whisper’s Transformer stack.
The final state of this token is used as the pooled audio representation:
\[
\mathbf{z}_a = \mathrm{WhisperEncoder}([\tilde H_0 ; \mathbf{a}_{\text{cls}}])_{[:, -1, :]}.
\]

Text inputs are encoded using either the \textit{CLIP text encoder} or \textit{ModernBERT},
yielding pooled text features $\mathbf{z}_t$ from its respective \textit{CLS} token.
Both modalities are mapped to a shared embedding space via lightweight
linear projectors, followed by $\ell_2$ normalization:
\[
\hat{\mathbf{u}}_a = \frac{f_a(\mathbf{z}_a)}{\|f_a(\mathbf{z}_a)\|_2},
\qquad
\hat{\mathbf{u}}_t = \frac{f_t(\mathbf{z}_t)}{\|f_t(\mathbf{z}_t)\|_2}.
\]

\subsection{Training Objectives}
\label{sec:method-obj}

We study two objectives.
\textit{CLIP loss (InfoNCE):}
\[
L = \tau \, \hat{\mathbf{u}}_a \hat{\mathbf{u}}_t^\top \in \mathbb{R}^{B \times B},
\]
where $\tau$ is a learnable temperature.
Cross-entropy is applied over rows (audio$\rightarrow$text) and columns (text$\rightarrow$audio).
\textit{SigLIP loss:} a sigmoid-based variant applying Binary Cross Entropy to all pairs,
labeling only diagonal pairs as positives.

\subsection{Training Strategies}
\label{sec:method-strategy}

We evaluate three adaptation regimes for the encoders:
(i) \emph{projector-only} (both encoders frozen),
(ii) \emph{LoRA} adapters in Transformer layers,
and (iii) \emph{full finetuning}.
In all cases, the global audio token is learned from scratch.

We also consider two update scopes:
(a) \emph{audio-only}, where the text tower is frozen, and
(b) \emph{both towers}, where audio and text are updated jointly.

\subsection{Design Sweep Configuration}
\label{sec:method-sweep}

Combining the factors of text encoder (CLIP vs.\ ModernBERT),
loss function (CLIP vs.\ SigLIP),
adaptation regime (3), and update scope (2),
we obtain a total of
$2 \times 2 \times 3 \times 2 = 24$ configurations.
These settings are explored in the design sweep
to study how different choices affect performance.

\subsection{Matryoshka Loss Adaptation}
\label{sec:method-matryoshka}

To enable dimensional scalability, we adopt Matryoshka supervision.
The idea is to train embeddings that remain useful
when truncated to smaller dimensions.
Let $d_1 > d_2 > \dots > d_K$ be target sizes
(e.g., $768 \!\rightarrow\! 384 \!\rightarrow\! 192 \!\rightarrow\! 96$),
and let $\mathrm{slice}(\cdot,d)$ select the first $d$ channels.
At each level $k$, contrastive loss is applied on the sliced embeddings:
\[
\hat{\mathbf{u}}^{(k)}_a = \mathrm{slice}(\hat{\mathbf{u}}_a, d_k), \quad
\hat{\mathbf{u}}^{(k)}_t = \mathrm{slice}(\hat{\mathbf{u}}_t, d_k).
\]
The overall loss is the mean across all levels:
\[
\mathcal{L} = \frac{1}{K} \sum_{k=1}^{K}
\mathcal{L}_{\text{contrast}}
\big(\hat{\mathbf{u}}^{(k)}_a, \hat{\mathbf{u}}^{(k)}_t\big).
\]

This produces a single model capable of emitting nested embeddings
at multiple scales.

\begin{table*}[t]
  \centering
  \vspace{-6pt}
  \caption{Retrieval performance with Recall@K on the AudioCaps and Clotho benchmarks. Shaded sub-rows ($\Delta$) show the performance change from using only the first 1/8 Matryoshka dimensions, with changes greater than 0.5 in magnitude marked {\color{mygreen}green for gains} and {\color{myred}red for losses}. \textbf{Top scores} are bolded; \underline{second-best} are underlined.}
  
  \label{tab:ret}
  \small
  \begin{tabular}{c ccc ccc ccc ccc}
    \toprule
    & \multicolumn{6}{c}{\textbf{AudioCaps}} & \multicolumn{6}{c}{\textbf{Clotho}} \\
    \cmidrule(lr){2-7} \cmidrule(lr){8-13}
    & \multicolumn{3}{c}{\textbf{Text-To-Audio}} & \multicolumn{3}{c}{\textbf{Audio-To-Text}} & \multicolumn{3}{c}{\textbf{Text-To-Audio}} & \multicolumn{3}{c}{\textbf{Audio-To-Text}} \\
    \cmidrule(lr){2-4} \cmidrule(lr){5-7} \cmidrule(lr){8-10} \cmidrule(lr){11-13}
    \textbf{Model} & 
    \textbf{R@1} & \textbf{R@5} & \textbf{R@10} & 
    \textbf{R@1} & \textbf{R@5} & \textbf{R@10} &
    \textbf{R@1} & \textbf{R@5} & \textbf{R@10} & 
    \textbf{R@1} & \textbf{R@5} & \textbf{R@10} \\
    \midrule
    
    
    

    WavLink-Large & \textbf{46.7} & \textbf{80.2} & \textbf{89.5} & \textbf{60.0} & \textbf{85.3} & \textbf{92.6} & \textbf{22.4} & \textbf{48.7} & \textbf{62.6} & \textbf{27.4} & \textbf{52.3} & \textbf{66.0} \\
    \rowcolor{gray!10}
    \hspace*{1em}\textit{$\Delta$ M-1/8} & \perfdelta{-0.3} & \perfdelta{0.1} & \perfdelta{0.1} & \perfdelta{-1.4} & \perfdelta{-0.3} & \perfdelta{0.0} & \perfdelta{-0.6} & \perfdelta{-0.2} & \perfdelta{-0.3} & \perfdelta{0.0} & \perfdelta{0.1} & \perfdelta{0.6} \\
    
    WavLink-Small & \underline{44.5} & \underline{79.0} & \underline{88.0} & \underline{54.3} & \underline{84.1} & \underline{92.3} & \underline{21.2} & \underline{46.8} & \underline{60.1} & \underline{25.3} & 49.4 & \underline{64.0} \\
    \rowcolor{gray!10}
    \hspace*{1em}\textit{$\Delta$ M-1/8} & \perfdelta{-0.6} & \perfdelta{-0.3} & \perfdelta{-0.1} & \perfdelta{0.2} & \perfdelta{0.4} & \perfdelta{0.4} & \perfdelta{0.2} & \perfdelta{-0.7} & \perfdelta{0.1} & \perfdelta{0.4} & \perfdelta{0.1} & \perfdelta{-1.1} \\
    
    WavLink-Base & 39.7 & 74.5 & 85.3 & 50.5 & 79.4 & 90.3 & 17.6 & 41.3 & 56.0 & 21.1 & 45.8 & 58.8 \\
    \rowcolor{gray!10}
    \hspace*{1em}\textit{$\Delta$ M-1/8} & \perfdelta{-0.1} & \perfdelta{-0.6} & \perfdelta{0.1} & \perfdelta{-0.6} & \perfdelta{0.5} & \perfdelta{-0.6} & \perfdelta{-0.5} & \perfdelta{0.0} & \perfdelta{-0.2} & \perfdelta{-1.0} & \perfdelta{-0.6} & \perfdelta{-0.4} \\

    \midrule
    \multicolumn{13}{c}{\textbf{Reported Performance From Prior Studies}} \\
    \midrule
    LAION-CLAP & 36.1 & 71.8 & 83.9 & 46.8 & 82.9 & 90.7 & 16.1 & 38.3 & 51.1 & 22.7 & 48.5 & 60.8  \\
    MGA-CLAP & 41.8 & 76.1 & - & \underline{54.4} & 83.6 & - & 20.4 & 46.0 & - & 25.3 & \underline{51.2} & -\\
    ReCLAP & 37.1 & 73.2 & 85.0 & 48.0 & 80.4 & 90.8 & 18.9 & 44.7 & 59.0 & 20.5 & 45.7 & 58.9  \\
    AF-CLAP & 37.3 & 72.9 & 84.0 & 46.9 & \underline{84.1} & 91.9 & 17.3 & 43.9 & 56.8 & 23.2 & \underline{51.2} & 63.5 \\

    \bottomrule
  \end{tabular}
\end{table*}

\begin{figure}[t]
\includegraphics[width=8.5cm]{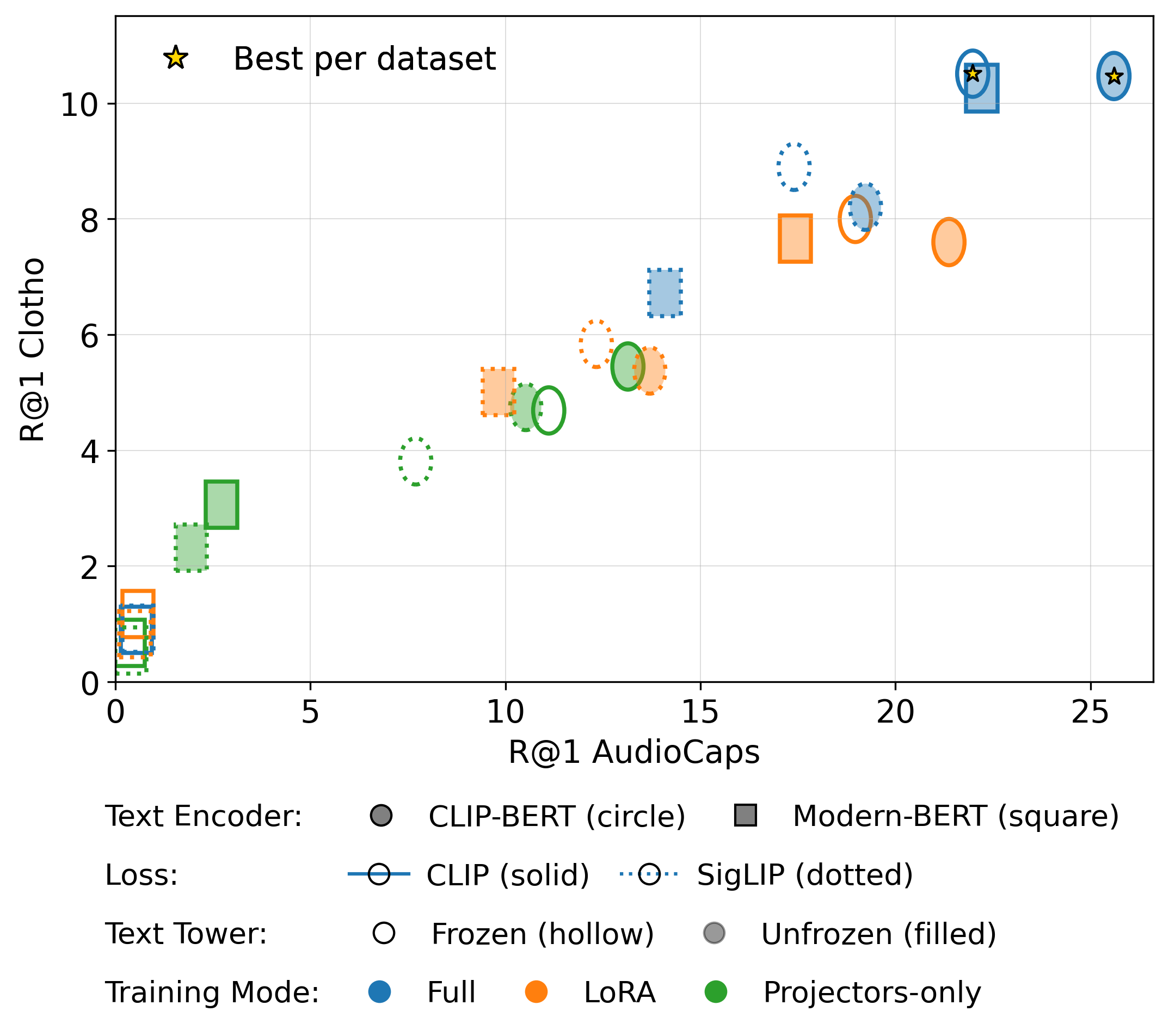}
\caption{Design sweep based on Recall@1 retrieval performance on the AudioCaps and Clotho benchmarks, identifying a fully trained CLIP-BERT model with CLIP's contrastive loss as the optimal configuration.}
\label{fig:design}
\end{figure}

\begin{table}[t]
  \centering
  \vspace{-6pt}
  \caption{Zero-shot classification performance with Accuracy (\%) on the VGG-Sound, US8K, and ESC-50 benchmarks.}
  \label{tab:class}
 \small 
  \begin{tabular}{c ccc}
    \toprule
    \textbf{Model} & \textbf{VGG-Sound} & \textbf{US8K} & \textbf{ESC-50} \\
    \midrule
    WavLink-Large & \textbf{31.7} & 74.5 & 83.0 \\
    WavLink-Small & \textbf{31.8} & 75.0 & 80.3 \\
    WavLink-Base  & 27.7 & 69.9 & 75.4 \\
    \midrule
    \multicolumn{4}{c}{\textbf{Reported Performance From Prior Studies}} \\
    \midrule
    LAION-CLAP & \underline{29.1} & 73.2 & 89.1 \\
    MGA-CLAP  & \textbf{31.8} & 83.7 & \textbf{94.9} \\
    ReCLAP  & \underline{29.2} & \textbf{95.2}  & \underline{92.8} \\
    AF-CLAP & 24.1 & \underline{92.3} & 91.3 \\
    \bottomrule
  \end{tabular}
\end{table}

\begin{table*}[t]
  \centering
    \vspace{-6pt}
  \caption{Multiple-choice QA performance with Accuracy (\%) on the AirBench Foundational  benchmark. Results for the Audio Encoder-LLM Decoder models are from ~\cite{falcon3audio}. LAION-CLAP is taken from \url{https://huggingface.co/laion/larger_clap_general}. \textbf{Top scores} are bolded; \underline{second-best} are underlined.}
  
  \label{tab:airbench}
 \small 
  \begin{tabular}{l cccc}
    \toprule
    & \multicolumn{2}{c}{\textbf{Dual Encoder}} & \multicolumn{2}{c}{\textbf{Audio Encoder-LLM Decoder}} \\
    \cmidrule(lr){2-3} \cmidrule(lr){4-5}
    
    \textbf{Task} & \textbf{WavLink-Base} & \textbf{LAION-CLAP} & \textbf{Qwen2-Audio Instruct} & \textbf{Falcon3-Audio 3B} \\
    \midrule
    
    $\sim$ parameter count in M (relative size)     & 84   & 193 (2x)  & 8400 (100x) & 3600 (43x) \\
    Number of audio tokens        & 1    & 1    & 750 & 750 \\
    \midrule
    
    \rowcolor{gray!10}
    \textbf{Total Average}                & \underline{42.0} & 35.8 & \textbf{44.0} & \underline{42.0} \\
    \midrule
    \rowcolor{gray!10}
    \textbf{Sound Average}                & 48.3 & 42.6 & \underline{49.8} & \textbf{53.4} \\
    Audio grounding              & 26.0 & 20.8 & 17.8 & 60.0 \\
    Vocal sound classification   & 71.3 & 55.9 & 71.1 & 74.9 \\
    Acoustic scene classification& 44.2 & 44.8 & 40.5 & 40.8 \\
    Sound question answering     & 50.8 & 43.5 & 62.8 & 52.4 \\
    \midrule
    \rowcolor{gray!10}
    \textbf{Music Average }               & \textbf{47.9} & \underline{46.2} & \underline{46.1} & 42.2 \\
    Music instruments classification & 62.5 & 56.2 & 49.6 & 46.9 \\
    Music genre classification   & 64.0 & 64.5 & 63.9 & 59.9 \\
    Music note analysis-pitch    & 26.0 & 31.4 & 24.3 & 19.6 \\
    Music note analysis-velocity & 27.8 & 25.6 & 24.7 & 22.8 \\
    Music question answering     & 33.1 & 29.5 & 56.0 & 41.8 \\
    Music emotion detection      & 40.7 & 38.6 & 38.7 & 39.9 \\
    \midrule
    \rowcolor{gray!10}
    \textbf{Speech Average }              & 34.4 & 24.7 & \textbf{43.5} & \underline{35.1} \\
    Speech grounding             & 29.2 & 25.4 & 26.3 & 20.3 \\
    Spoken language identification & 27.1 & 19.3 & 38.1 & 4.9 \\
    Speaker gender recognition   & 53.3 & 35.9 & 52.5 & 23.9 \\
    Speaker emotion recognition  & 37.3 & 19.7 & 35.4 & 60.3 \\
    Speaker age prediction       & 13.5 & 25.3 & 22.3 & 26.4 \\
    Speaker entity recognition   & 25.3 & 22.5 & 48.3 & 26.2 \\
    Speaker intent classification& 31.4 & 27.0 & 78.0 & 51.4 \\
    Speaker number verification  & 54.8 & 21.4 & 38.0 & 39.2 \\
    Synthesized voice detection  & 16.2 & 20.1 & 51.9 & 49.3 \\
    \bottomrule
  \end{tabular}
\end{table*}

\section{Experiments}
\label{sec:exp}

\subsection{Experimental Setup}
\label{sec:exp-setup}

\textbf{Datasets.}
The design sweep is trained on $\sim$2M audio--text pairs from Auto-ACD
(AudioSet + VGGSound derived).
Scaled training uses two stages:
Stage-1 with additional $\sim$6M captions from AudioSetCaps (AudioSet + VGGSound + YouTube-8M derived),
and Stage-2 with $\sim$0.1M captions from AudioCaps v2 and Clotho training splits.

\textbf{Models.}
\emph{Audio encoders Initialization:} Whisper-Large(v3) for \textit{Large}, Whisper-Small(en) for \textit{Small}, Whisper-Base(en) for \textit{Base}.
\emph{Text encoders Initialization:} CLIP-ViT-L/14 or ModernBERT-Large for \textit{Large}, CLIP-ViT-B/32 for \textit{Small} and \textit{Base}.
\emph{Projectors}: Single linear layer to match the projector dimension, with pretrained weights used for CLIP based text projectors. The global token is trained from scratch always. More model specifications are shown in Table~\ref{tab:model_specs}.

\textbf{Training.}
\emph{Framework:} PyTorch Lightning.
\emph{Common configuration:} DDP strategy; BF16 precision; AdamW optimizer; 1e-4 learning rate; cosine scheduler with 5\% warmup; Embeddings are gathered across GPUs before computing CLIP loss.
\emph{Design sweep configuration:} 8$\times$H100 80GB GPUs; 80 batch size;
 10 epochs; \textit{Large} variant; LoRA rank 8. 
\emph{Scaled runs configuration:} 64$\times$H100 80GB GPUs; 768 batch size; 3 epochs per stage; Matryoshka supervision with $K=4$ and dimensions ${d, d/2, d/4,}$ and ${d/8}$.


\subsection{Design Sweep Results}
\label{sec:exp-sweep}

We evaluate 24 setups
(2 encoders $\times$ 2 losses $\times$ 3 regimes $\times$ 2 scopes).
Figure~\ref{fig:design} shows R@1 on AudioCaps and Clotho.
The best setting is \emph{CLIP text, CLIP loss, full finetuning, both towers updated},
adopted for scaled Stage-1/2 training.
Interestingly, ModernBERT underperformed CLIP despite its stronger text
benchmarks, suggesting that CLIP’s alignment priors transfer better to audio--text retrieval.
\subsection{Retrieval Performance}
\label{sec:exp-retrieval}

Table~\ref{tab:ret} reports results on AudioCaps and Clotho.
\textit{WavLink-Large} improves over CLAP baselines by $\sim$2--6 points across R@K.
\textit{WavLink-Small} trails by only 1--2 points while using $\sim$20\% of parameters.
\textit{WavLink-Base} (<100M parameters) is competitive with existing CLAP based models showing that Whisper-derived embeddings can rival models explicitly trained for audio--text alignment.

\subsection{Generalization Beyond Retrieval: ZSC}
\label{sec:exp-gen-zsc}

\textit{WavLink} achieves top zero-shot classification accuracy (Table~\ref{tab:class}) on VGGSound.
On ESC-50 and US8K it lags behind prior models, potentially due to relatively more dense training captions.
Task-specific fine-tuning could close this gap.

\subsection{Generalization Beyond Retrieval: MCQ}
We reframe multiple-choice QA as zero-shot classification and show the performance on AirBench Foundational. 
We combine the question with each candidate choice, and pick the option whose joint text embedding is most similar to that of the audio.
As shown in Table~\ref{tab:airbench},
\textit{WavLink-Base} (84M, 1 token) achieves 42.0\%,
+6 over LAION-CLAP and comparable to Falcon3-Audio-3B,
while trailing Qwen2-Audio Instruct by only 2 points
despite being 43--100$\times$ smaller.
Performance is strong on classification tasks across speech, sound, and music,
but weaker on grounding and fine-grained analysis.
This pattern is intuitive: speech-heavy tasks benefit from Whisper’s ASR pretraining,
while grounding requires finer token-level alignment that frame-based LLM methods capture better.

\subsection{Scalability }
\label{sec:exp-scale}

Embeddings compressed to 1/8 dimension maintain accuracy
within $<$1 point on average (Table~\ref{tab:ret}),
reducing storage and similarity compute by 8$\times$.
Larger models degrade less under compression, suggesting redundancy.
This property is especially valuable for web-scale retrieval,
where both storage and similarity search cost are dominant bottlenecks.

\subsection{Ablations}
\label{sec:ablations}

Replacing the Whisper encoder with a pretrained HTS-AT from LAION-CLAP in the \textit{large} setup yielded R@1 (T2A/A2T) scores of 45.8/56.4 on AudioCaps and 14.0/14.6 on Clotho. While lower than \textit{WavLink-Large} on AudioCaps, the drop was more severe on Clotho, where it underperformed even \textit{WavLink-Base}. This indicates HTS-AT’s limitations with longer audio segments ($>$10s, common in Clotho), underscoring Whisper’s robustness as a backbone for general-purpose audio--text embeddings.

\section{Conclusion}
\label{sec:con}

We introduced \textit{WavLink}, a compact audio--text embedding model
that augments Whisper with a learnable global token.
Through systematic design sweeps, scaled two-stage training,
and Matryoshka supervision, WavLink achieves
state-of-the-art retrieval performance,
strong zero-shot classification on VGGSound,
and competitive results on AIR-Bench
while using a single global token instead of 1500 frame-level features.
These findings highlight the underexplored potential of Whisper
beyond speech recognition for efficient representation learning.

Future directions include extending WavLink to multilingual
audio--text alignment and leveraging the global token mechanism
for audio--LLMs, where compact and adaptive embeddings
can reduce compute cost while improving cross-task generalization.

\clearpage
\bibliographystyle{IEEEbib}
\bibliography{refs_short}

\end{document}